\documentclass[aps,twocolumn]{revtex4}
\usepackage{graphicx}
\usepackage{amsmath}
\usepackage{latexsym}
\usepackage{amsmath}
\usepackage{amssymb}
\usepackage{float}
\usepackage{epstopdf}
\usepackage{hyperref}
\hypersetup{urlcolor=red,citecolor=blue}
\usepackage{bm}
\begin{document}
\title{A synergistic view of magnetism, chemical activation, and ORR as well as OER catalysis of 
carbon doped hexagonal boron nitride from first-principles}
\author{Rita Maji and Joydeep Bhattacharjee\\
\small{\textit{School of Physical Sciences}}\\
\small{\textit{National Institute of Science Education and Research}}\\
\small{\textit{HBNI, Jatni - 752050, Odisha, India}}}
\begin{abstract}
Carbon(C) doped hexagonal boron nitride(hBN) has been experimentally reported in recent years to be a possible catalytic host to 
oxygen reduction reaction(ORR), as well as a possible ferromagnet at room temperature.
Substitution by C in hBN has been also reported to form islands of graphene. 
In this work, we explore from first principles, the connection between these different aspects of C doped hBN.
We find formation of graphene islands covering unequal number of B and N sites in hBN to be energetically plausible. 
They posses a net non-zero magnetic moment and are also found to be substantially more chemically active than their non-magnetic
counterparts covering equal number of B and N sites.
On-site Coulomb repulsion between electrons, known to be responsible for magnetism in bipartite lattices like graphene and hBN,
is also found to play a central role in chemical activation of not only the C atoms at the zigzag interface of 
magnetic graphene islands and hBN, but also of boron(B) sites in the immediate hBN neighborhood.
However, such activated B or C due to substitution at B site, which is energetically more favorable than at N site, 
has been reported to be unfavorable for ORR. 
Advantageously, we find that the activation of C at B sites moderates systematically with increasing size of graphene islands,
paving the way for abundance of efficient catalytic sites at the edges of magnetic graphene islands covering more B sites than N sites.
Accordingly, as an alternate to precious metals for electrodes, we propose a class of graphene-hBN hybrids with lattices of 
magnetic graphene islands embedded in hBN, which can be metallic.  
    
\end{abstract}
\maketitle
\section{Introduction}
Boron(B) and/or nitrogen(N) doped graphene has been extensively  
explored\cite{rev-gra-mat-orr2013,rev-gra-mat-catal2014,rev-B-N-gra-orr2014,orr-N-gra-dft2015} in the last decade or so, 
primarily in pursuit of an efficient metal free catalytic platform for oxygen reduction (ORR)\cite{rev-gra-mat-orr2013},
and evolution reaction (OER)\cite{gnf-oer2016,N-gnf-oer2016}.
These efforts are aimed broadly at replacing the expensive metal based cathodes in fuel-cells\cite{rev-metalfree-orr2015}.
Conversely to B and N doped graphene, carbon(C) doped hexagonal boron nitride(hBN) 
has also come under focus in recent years as catalytic host of ORR\cite{orr-C-hBN-dft2015,orr-gra-hbn-2016,orr-CatB-hbn-2016}, 
since the neighborhood of an active C in B and N co-doped\cite{orr-BCNgra2012,rev-B-N-gra-orr2014} graphene
is expected to be similar to that of a carbon(C) atom substituting a B or N atom in hBN.
However, computational studies in this direction so far are  based on single active C sites in hBN, 
whereas experiments in last few years tend to suggest that substitution by C in hBN  
might prefer formation of islands of graphene in hBN\cite{doped-C-patch-hbn2013-B, syn-gra-hbn-domain2013,doped-C-patch-hbn2013-A}. 
C doped hBN has been also in focus in recent years for a host of possible device applications\cite{hbn-gr-application}, owing to their
tunable electronic and magnetic properties primarily due to the confinement of the 2$p_z$ electrons in islands of graphene
and the nature of Gr-hBN interface. 
They have been experimentally reported to be ferromagnetic at room temperature in recent years\cite{expt-fm-BN-GR}.
In this work our aim is to take a synergistic view of all these aspects of C doped hBN to correlate
magnetism with chemical activation and catalysis of C doped hBN.

Substitution by C is energetically more favorable at B sites than that at N sites because the C-N bonds are lower in energy, 
than the C-B bonds, owing to the ascending order of electronegativity of B, C and N.
The energy of the extra(remaining)  2$p_z$ electron at B(N) site due to substitution by C must by close the conduction(valence) band edge 
but a bit lower(higher) due to one extra(less) proton at C compared to B(N).
Therefore, a  2$p_z$ electron of C at B site will have a higher energy than its counterpart at N site.
However, substitution of two C atoms from neighboring sites should allow their 2$p_z$ electrons to delocalize and form $\pi$ bond 
to reduce their kinetic energies.
Interplay of these factors will thus determine the energetics of substitution by C in hBN.
Notably, a C atom at a B or N site will not be able to complete its sub-shell filling, and will thereby be chemically active,  
unless has another C atom in its nearest neighborhood free to make a $\pi$ bond with it.
 Furthermore, the spin carried by the unpaired 2$p_z$ electrons of C atoms substituting in hBN are source of magnetism of 
C doped hBN. These unpaired electrons can thus either belong to an isolated C atom due to a single substitution,
or to a graphene island which would not allow all C atoms to complete sub-shell filling.
Chemical activation and magnetism of C doped hBN are thus inherently related.

Chemically active sites are catalytically active selectively depending on the reaction.
Isolated C at N(B) site has been shown to be favorable(unfavorable) as catalytic support to oxygen reduction reaction (ORR)
owing to over binding of -OOH, -OH and -O intermediates\cite{orr-C-hBN-dft2015} in the ORR pathway.
ORR on such sites thus does not remain completely spontaneous even after exhausting all the 
energy available at cathode due to formation of water.
Rather, on C at N sites, ORR pathway has been shown to remains spontaneous all through out without exhausting all the energy
available at cathode, which leaves some energy to drive electron through external load, implying non-zero operating voltage. 
Since substitution by C at B site is more favorable than that at N site, 
the ORR active sites will thus be less available than the inactive ones.
However, since such single C sites are not likely to sustain due to preference of successive substitution by C at neighboring sites, 
ORR activation of C doped hBN needs to be revisited in the light of formation of islands of graphene in hBN.
We show in this work that with increasing island size the activation of C atoms weakens in general leading to
enhanced availability of ORR active sites.

\begin{figure}[t]
\centering
\includegraphics[scale=0.19]{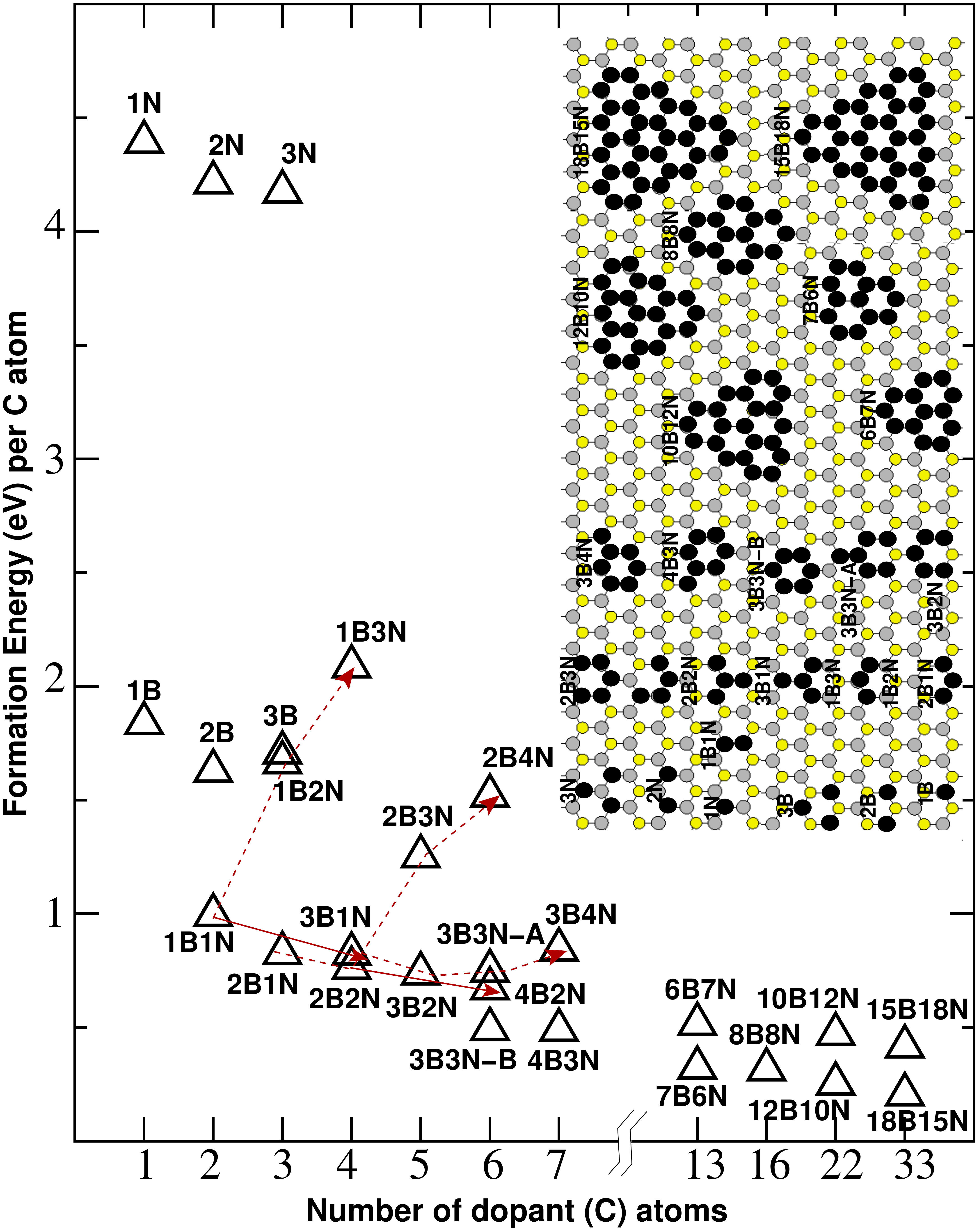}
\caption{Average formation energies (per C atoms) for a representative variety of configurations of substitution 
by C in hBN shown in inset. 
Dashed(solid) arrow indicates the variation in average formation energy with increasing number of substitutions at N(B) sites for
fixed numbers of B(N) sites.}     
\label{Eform}
\end{figure}

To rationalize the plausibility of formation of chemically active graphene islands in hBN, 
we first compute\cite{method-dft} the energetics of substitution by C in a representative variety of configurations 
with increasing number of C atoms, as shown in Fig.\ref{Eform}. We have considered triangular graphene islands 
to represent graphene islands with non-zero magnetic moment and studied their chemical activation in conjunction with their magnetism.
To understand the microscopic origin of activation, we have constructed spatially localized Wannier functions\cite{method-WF,orr-N-gra-dft2015} 
which reveal an orbital resolved mechanism highlighting the role of on-site Coulomb repulsion in chemical activation.  
Over-potential required to complete ORR and OER pathways on active sites are estimated by calculating Gibb's free energies of reaction steps.
Substantial lowering of over-potential is found for C atom at the zigzag interfaces around magnetic graphene islands covering more B sites, 
compared to a single C atom at B site.
Accordingly, we suggest a super-lattice of magnetic graphene islands in hBN as a cathode(anode) for ORR(OER) with acidic electrolyte.
\section{Computational details}
Total energies are obtained using a plane wave based implementation of DFT \cite{method-qe} where the
ionic potentials are approximated by $ultrasoft$\cite{method-usp} pseudo-potentials 
which are maximally smooth in the core region of atoms.
Exchange-correlation contribution to total energy is estimated approximately using a gradient corrected 
Perdew-Burke-Ernzerhof (PBE) \cite{method-pbe} functional. 
Configurations of substitutions by C, namely, islands of graphene, are considered in large hBN super-cell 
such that the islands are at least 10 \AA{} apart in their closest approach. 
Minimum energy configuration are
obtained using the BFGS \cite{method-bfgs} scheme of minimization of total energies, which are
converged with plane-wave cutoff more than 800 eV, k-mesh equivalent to 35$\times$35 per primitive unit-cell of pristine hBN, 
and forces less than $10^{-4}$ Rydberg/Bohr for all atoms.
Chemisorbed configurations are obtained within PBE, which is known to underestimate binding energy and thereby adsorption as well. 
Physisorbed configurations have been further relaxed by incorporating dispersion interaction\cite{grimme}.

To derive orbital resolved mechanism of chemical activation we construct Wannier functions(WFs) which can be understood 
as linear combinations of Bloch functions, which are obtained from
first principles as K-S states using DFT. In one dimension the set of WFs which have maximum localization in a given direction, 
also exclusively diagonalizes the first moment (position operator) along the given direction in the basis of occupied states. 
Therefore, for the isolated system, such WFs can be readily constructed by diagonalization of position operator for the given direction 
(e.g. $\hat{x}$) in the occupied K-S basis $\{\phi_m^{KS}\}$ obtained as,
\begin{equation}
 X_{mn}=\langle\phi_m^{KS}|\hat{x}|\phi_n^{KS}\rangle \nonumber
\end{equation}
For periodic systems, $\underline{\underline{X}}$ can be computed from geometric phases.
However, since the position operators along three linearly independent directions ($X, Y, Z$),
may not commute within a finite basis set, they can be approximately joint diagonalized to obtain a set of WFs,
which maximally diagonalizes all three their first moment matrices simultaneously and are thereby localized in all three dimensions. 
These WFs, unlike the maximally localized Wannier 
function\cite{mlWF}, does not depend on any reference template of orbitals to define their 
symmetries. In this method, the Wannier centers(WCs), which are basically the center of mass of WFs, can be obtained directly from the approximate 
eigenvalues of the three first moment matrices(FMMs), without explicitly constructing the WFs.   

Each WC for a given spin represents one electron, facilitating precise estimation of number of electrons associated 
with bonds and atoms. The position of WCs provides a unique dot structure map for the valence electrons over the entire system cell.
Based on the location of WCs with respect to atoms per spin, WCs can be segregated in two categories:(1) WCs associated with atoms and (2)
WCs along the bonds connecting two atoms respectively.
Single and double bonds are represented by one and two WCs between two nearest neighboring atoms.

To estimate the number of valence electrons of a given atom, we consider one electron for each of the WCs associated with the atom, and 
half an electron for each of the WCs along all the bonds made by the atom. 
Notably, this counting process is markedly different from any other localized descriptions like Mulliken, Bader or L\"{o}wdin analysis. 
The level of sub-shell filling of a given atom is estimated by consider one electron each for all the WCs associated with the atom 
as well as the bonds made by the atom.
In this work we have constructed isolated hydrogen passivated segments to construct WCs.
\begin{figure*}[t]
\centering
\includegraphics[scale=0.33]{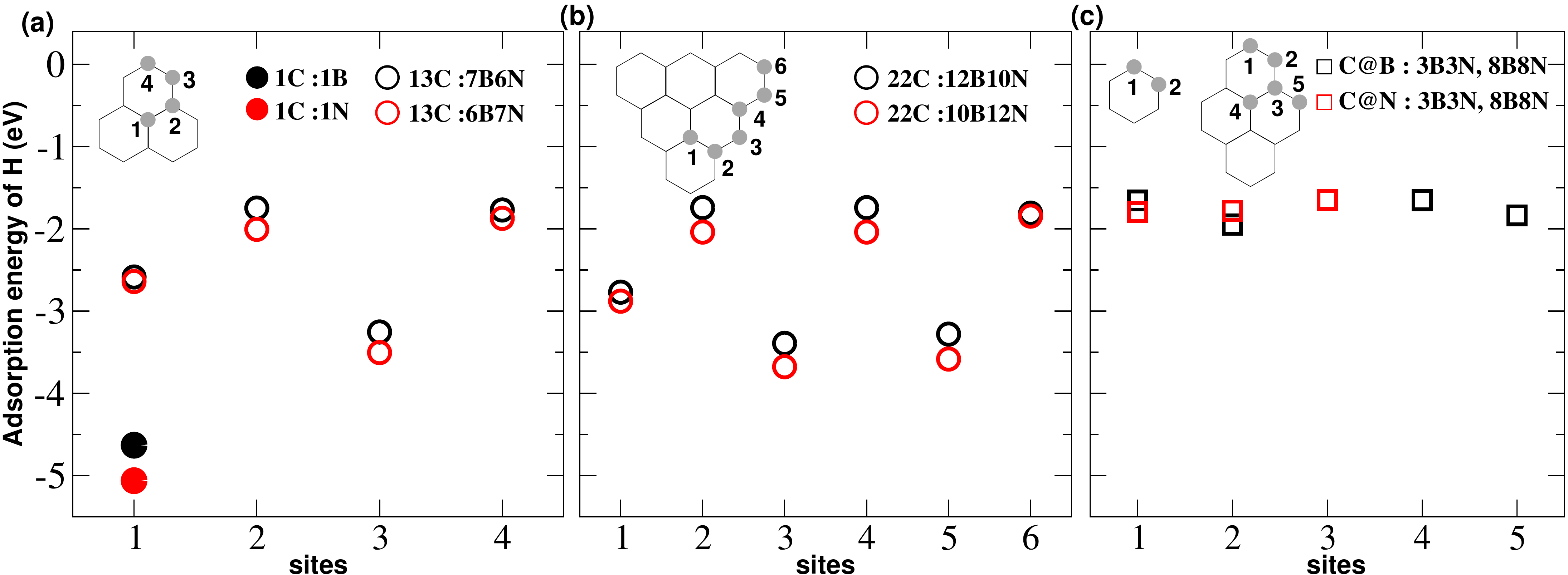}
\caption{Atomic H adsorption energies on C atoms marked by gray circles in (a)C13, (b)C22 and (c) C6 and C16.
Note that most of the C atoms considered, are inequivalent.
Circles marked in red(black) are C at N(B) site. Solid circles in (a) represent adsorption energies on isolated C sites in hBN.}     
\label{Hadsorption}
\end{figure*}
\section{Results and discussion}
we consider a representative variety of configurations
of substitution by C in hBN, with an increasing number of C atoms as described in
Fig.\ref{Eform}. A general nomenclature of (n$_B$ +n$_N$)C=n$_B$ B + n$_N$ N has been used to denote the 
number of B(n$_B$) and N(n$_N$) sites substituted by C in the immediate neighborhood of
each other. Thus, n$_B$ = 0 or n$_N$ = 0 would mean substitution only from next nearest
sites constituting the N or B sublattices, whereas, n$_B$ = n$_N$ would imply substitution
from nearest neighboring sites such that all the C atoms can in principle have a double
bond in pairs. By the same argument, n$_B$ = n$_N$ would imply |n$_B$ − n$_N$| number of
C atoms to have all three single bonds and would thereby have incomplete sub-shell
filling unless negatively charged. It is the ratio of number of C-C single and double
bonds(C=C) and the distribution of C atoms with all single bonds, which determine
the energetics of substitution.
\subsection{Energetics of substitution} 
Average(per C atom) formation energy(AFE)\cite{postsynthesis2011} of a C doping configuration due to substitution at 
$n_B$ and $n_N$ numbers of B and N sites
respectively, is estimated in N$_2$ rich condition as:
\begin{equation}
\left\{(E-E_0)+ n_N\mu_N+n_B \mu_B-(n_N+n_B)\mu_C\right\}/(n_N+n_B), \label{efm}
\end{equation}
where $E$ and $E_0$ are total energies of a large enough (8x8) hBN super-cell with and without substitution by C. 
$\mu_C$ and $\mu_N$ are chemical potentials which are estimated as energy per atom in graphene and N$_2$ respectively.
$\mu_B$ is calculated as $\mu_{hBN}-\mu_N$ where $\mu_{hBN}$ is approximated as total energy per B-N pair (primitive cell) in hBN.
AFEs[Fig.\ref{Eform}] suggest that energetics of substitution is determined by the competing energetics of the 
$\sigma$ and $\pi$ electrons of the C atoms.
Energetics of the $\pi$ electrons are governed by the interplay of delocalization in attempt to reduce kinetic energy,
and localization due to Coulomb repulsion. For example, the sharp drop in AFE for 1B1N is due to delocalization of $\pi$ electrons 
along C-C bond.
Successive substitution by C would will be preferred at neighboring pairs of sites, such that the compressive strain caused by the
C=C (double) bonds, owing to their shorter length compared to the B-N bonds, can be compensated by the modest expansive strain due to 
C-C single bonds, which are longer than the B-N bonds but not as long as the C-B bonds.
Furthermore, with $n_B=n_N=3$, Fig.\ref{Eform} suggests that substitution by C in closed loop 
is preferable over an open chain, since an open chains would leave uncompensated  strain fields in hBN at the
terminal C atoms. Therefore, successive substitution by C in hBN is preferred at nearest neighboring sites in closed loops.
Thus in formation of bigger islands of graphene in hBN, 
not only the higher energy cost of substituting atoms is increasingly compensated by lowering of kinetic energy of the $\pi$ 
electrons due to increased $\pi$-conjugation, but strain is also reduced at the periphery of the islands.
Comparison of AFEs for 
bigger islands with $n_B=n_N$ and $n_B\ne n_N$ suggests that islands of both types will be accessible increasingly on equal footing 
with increasing islands size.  
\subsection{Chemical activation}
Notably, the bigger islands C13 (7B6N,6B7N), C22 (12B10N,10B12N) and C33 (18B15N,15B18N) would have a total of 1, 2 and 3 
unpaired 2$p_z$ electrons respectively, since the rest of them can be accounted for in 6, 10 and 15 $\pi$ bonds 
between distinct pairs of  nearest neighboring C atoms. 
Those islands would therefore have 1,2 and 3 active C atoms respectively, each with incomplete sub-shell filling. 
However, since the unpaired 2$p_z$ electrons will delocalize across the island due to resonating $\pi$ conjugation 
configurations, activation will also be distributed symmetrically across the island. 
The trend that in C13, C22 and C33 islands, 1, 2 and 3 unpaired electrons are distributed across 13, 22 and 33 sites respectively, 
hints at possible moderation of activation as we go from 
C1(1B or 1N) to larger islands. 
To look for any such trend, we estimate adsorption energy[Fig.\ref{Hadsorption}] of atomic hydrogen(H) on the C1 and then also
on the inequivalent C atoms of C6, C13, C16 and C22.
We find the adsorption is strongest on 1N, followed closely by 1B, weakening further on the zigzag interfaces of C13 followed by C22, 
and the weakest on C6 and C16. 
In the islands, the most active sites are symmetrically located on and about the middle of the zigzag interfaces.
These trends thus not only confirms the anticipated trend of lowering of activation
of C atoms with increasing size of graphene islands in hBN, but also suggests that for activation is weak and uniformly distributed
if $n_B=n_N$, which can be understood by noting that in those islands all the C atoms can complete sub-shell filling on equal footing.
We note here that the level of chemical activation of a site can vary with variation in reactants. 
Therefore our discussion of activation is always with respect to a certain reactant, which is atomic H in this paper, 
unless stated explicitly. 
\subsubsection{Role of magnetic order in activation}
\begin{figure}[t]
\centering
\includegraphics[scale=0.4]{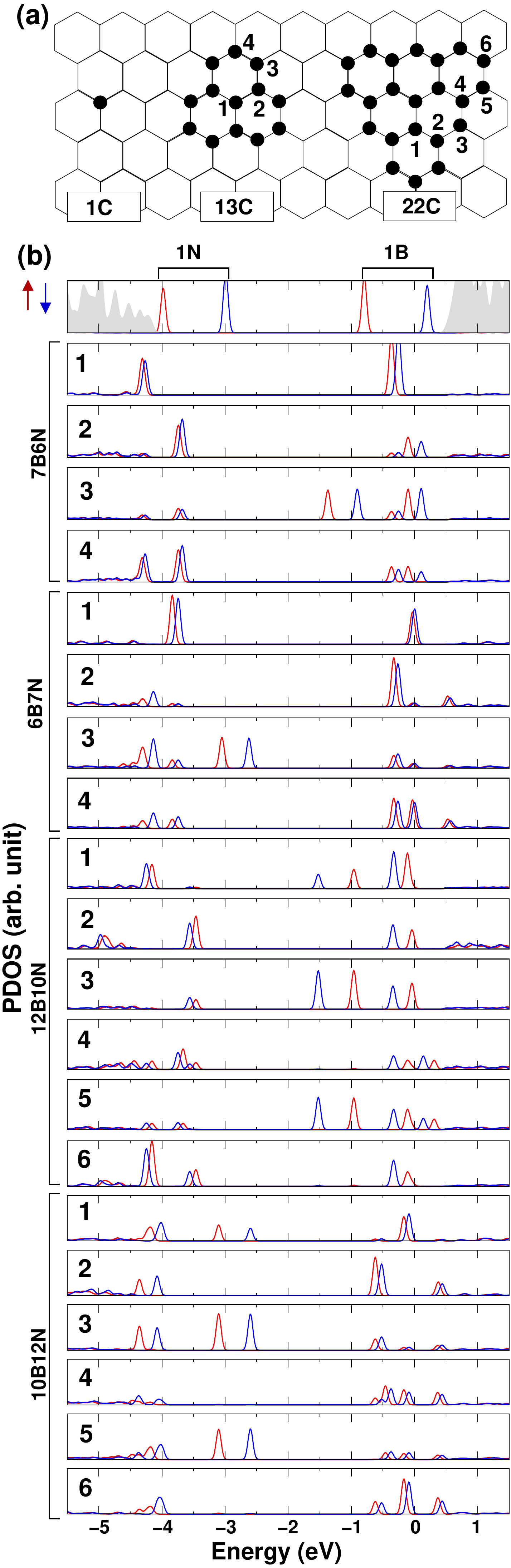}
\caption{(a)C1, C13 and C22 islands with C atoms marked. 
(b) Density of states projected on 2p$_z$ orbitals of the marked C atoms in C1(1N,1B), C13(7B6N,6B7N) and C22(10B12N, 12B10N) islands.}     
\label{PDOS}
\end{figure}
\begin{figure}[t]
\centering
\includegraphics[scale=0.12]{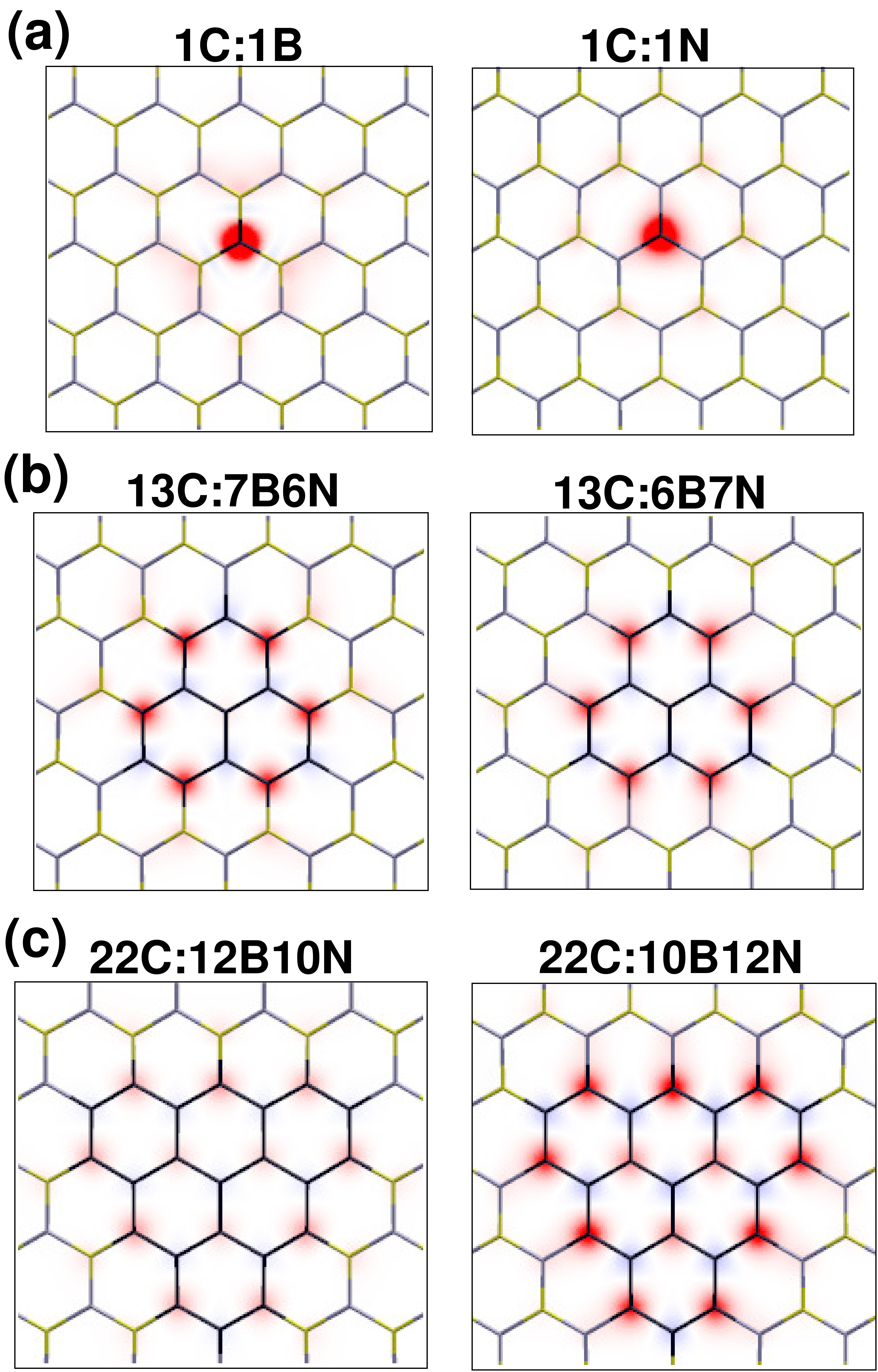}
\caption{Spin densities of (a)C1, (b)C13 and (c) C22 islands.}     
\label{spindensity}
\end{figure}
As per the Lieb's theorem for bipartite lattices, C13, C22 and C33 islands would posses net magnetic moments respectively of 
1, 2 and 3 $\mu$B, since the coverage of sites in the N and B sublattices by the islands differ by 1, 2 and 3 
respectively. C6 and C16 will therefore have no net magnetic moment. 
Therefore, as per Fig.\ref{Hadsorption}, graphene islands with net nonzero magnetic moment are chemically more active than the 
ones with no magnetic moment, implying that magnetism and chemical activation are inherently intertwined. 
We notice in Fig.\ref{Hadsorption} that the activation among the inequivalent C atoms maximises at the zigzag edges.
To understand this trend we plot[Fig.\ref{PDOS}] the density of states(PDOS) projected on the 2$p_z$ orbitals of the 
inequivalent C atoms in the C13 and C22 islands and compare them with those of a single isolated C atom. 
For 7B6N (6B7N), Fig.\ref{PDOS} implies that six equivalent carbon atoms at B (N) sites at the zigzag interface, 
one of which is marked as 3 in 13C in 
Fig.\ref{PDOS}(a), have not only their occupied 2$p_z$ orbitals lowered (raised) in energy compared to the other inequivalent C atoms at
B (N) sites, but also have larger splitting  
in energy of spin polarized 2$p_z$ orbitals: $\Delta E(2p_z)=|E(2p_z)_{\uparrow}-E(2p_z)_{\downarrow}|$. 
Increased $\Delta E(2p_z)$ about Fermi energy implies increased on-site Coulomb repulsion due to strong
localization of spin polarized 2$p_z$ orbitals.
This is readily confirmed by the spin density plotted in Fig.\ref{spindensity}, where we see strong spin polarization 
of sites with maximum chemical activation as determined in Fig.\ref{Hadsorption}.
Fig.\ref{spindensity} thus clearly suggests that the ascending order of activation follows the ascending degree of spin polarization. 
For example, among C at B sites, the isolated substitution C1:1B results into the most active C atom which 
has the highest $\Delta E(2p_z)$ and thus the largest spin density.
Whereas, the site 1 in C13:7B6N, which has its occupied 2$p_z$ orbital higher in energy than C1, but has minimal $\Delta E(2p_z)$, 
is the least active.
A C atom in a B site rich island like 7B6N, and its equivalent counterpart in an N site rich island of similar size like 7N6B,
have similar $\Delta E(2p_z)$ [Fig.\ref{PDOS}], but have different energies of the occupied 2$p_z$ orbitals, 
would differ marginally in their chemical activation. 
Among the inequivalent C atoms at the same sublattice (B or N), the variation in activation is stronger and determined 
exclusively by $\Delta E(2p_z)$.
Thus the $\Delta E(2p_z)$ plays the role of the dominant quantitative marker for comparison of chemical activation of 
sites in the same sublattice.
\subsubsection{Orbital resolved picture of chemical activation}
\begin{figure}[b]
\centering
\includegraphics[scale=0.37]{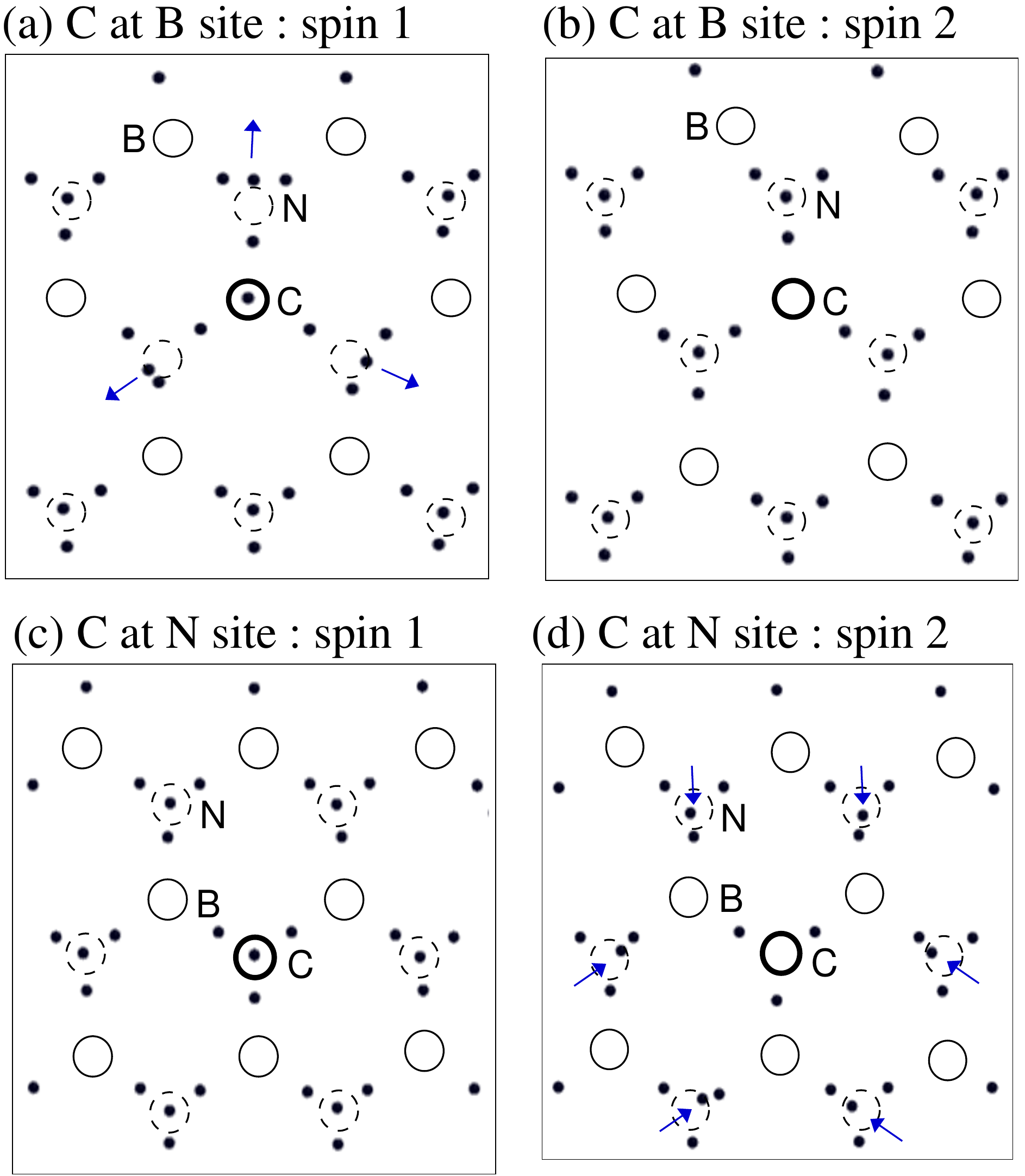}
\caption{Wannier Centers(WCs) shown as black dots for (a)spin1, (b)spin2 with C at B, and (c)spin1, (d)spin2 with C at N.
The blue arrows indicate shift of WCs after substitution by C, compared to WCs in pristine hBN.
Solid(dashed) circles indicate B and N atoms. }
\label{WC1B1N}
\end{figure}

Spin resolved WCs plotted in Fig.\ref{WC1B1N}(a,b and c,d) for 1B and 1N respectively,
clearly suggests attempts to create spin separation between the two sublattices in the vicinity of C atom
under the action of the on-site Coulomb repulsion.
The arrows in Fig.\ref{WC1B1N}(a and d) indicate shift of the centre of mass of only one of the two 2$p_z$ orbital of N, namely,
the one which has the same (opposite) spin as(to) that of the 2$p_z$ orbital of the nearest(next-nearest) neighboring C atom in 1B(1N). 
Such spin separation is completely suppressed in pristine graphene by the lowering of kinetic energy due to $\pi$ conjugation.
In hBN, the large difference in electronegativity of B and N suppresses both $\pi$ conjugation as well as spin separation 
and results into wide band-gap.   
However, at the zigzag interface of graphene and hBN hindered $\pi$ conjugation among C atoms allow Coulomb correlation to 
consolidate and determine physical and chemical properties of the interface, as we see here.
Notably, the displacement of WCs imply increase in order of the B-N bonds
and deviation from charge neutrality and sub-shell filling of B and N atoms.
These deviations, which leads to increase in total energy, can be removed by engaging the unpaired electron in chemisorption on the active C atom.
Thus, besides playing a direct role in chemical activation, on-site Coulomb repulsion also plays an in-direct role 
in facilitating chemisorption on the active C atoms as a means to preserve the electronic structure of the hBN neighborhood.

\begin{figure}[t]
\centering
\includegraphics[scale=0.1]{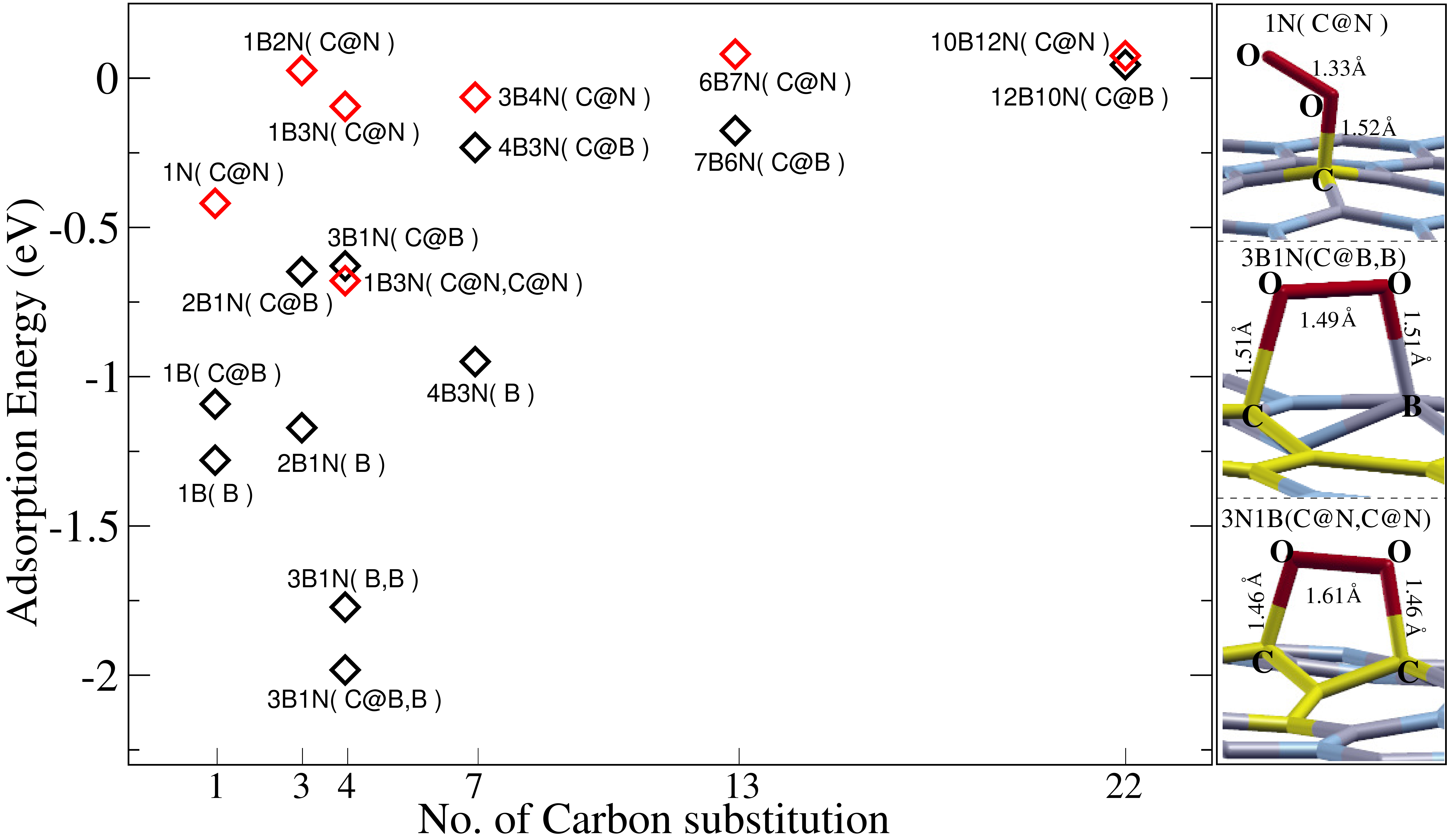}
\caption{Adsorption energy of O$_2$ on a representative variety of configuration of substitution by C in hBN 
starting from isolated cite to bigger islands. 
Values marked in red(black) are for adsorption on C at N(B) site on N(B) site rich islands. 
For islands (C13,C22) the adsorption sites are on the zigzag interface of Gr and hBN. 
For smaller islands the adsorption sites are the outermost ones.}
\label{O2adsorption}
\end{figure}
\begin{figure}[b]
\centering
\includegraphics[scale=0.13]{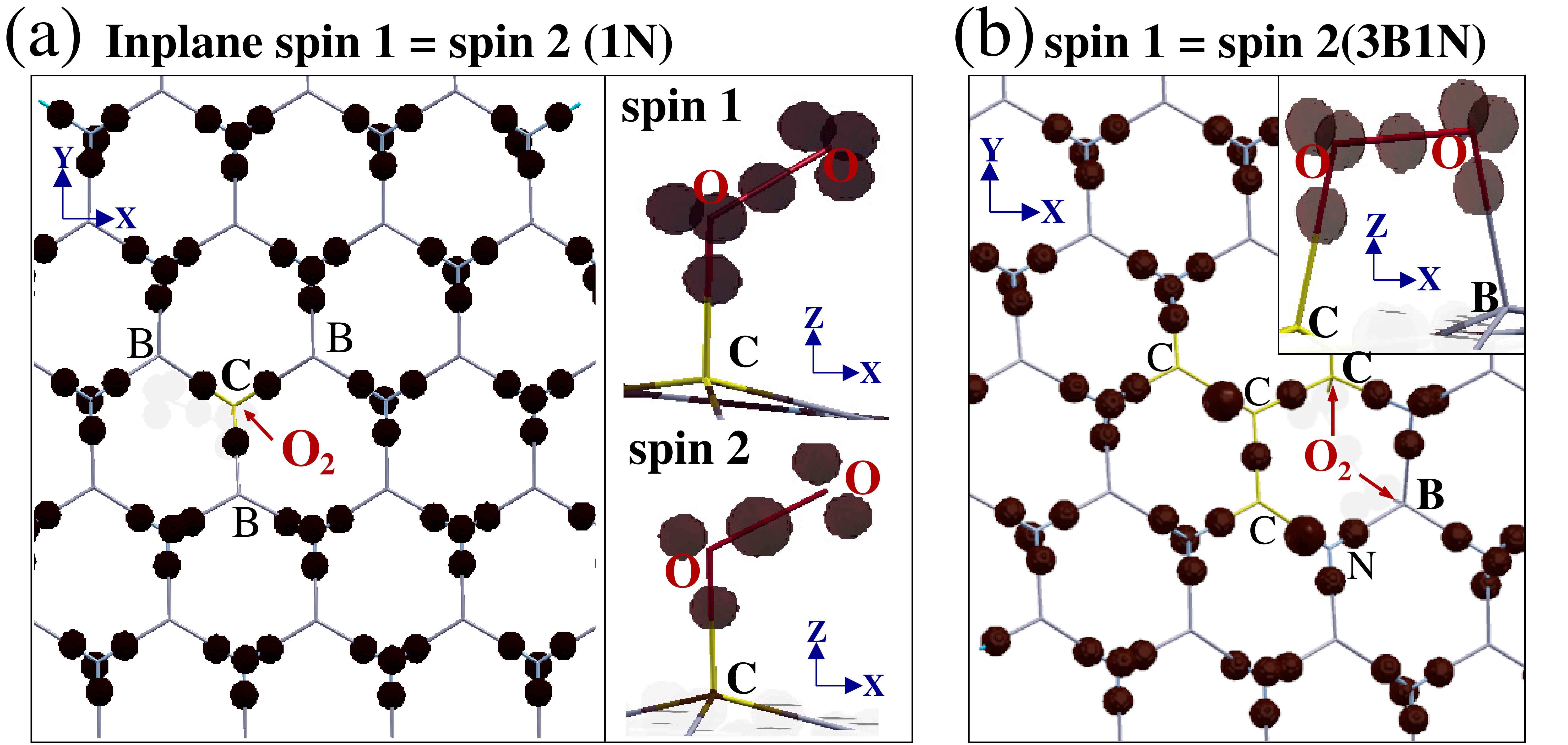}
\caption{Wannier Centers(WCs) upon adsorption of (a)O$_2$ as semi-diradical, (b)O$_2$ as  neutral bridge, where small(big) dark translucent
spheres are used to represent  one(two) electron(s).}
\label{WCO2adsorption}
\end{figure}
\subsection{Adsorption of O$_2$}
Within LSDA with PBE(GGA) we find O$_2$ to carry a net magnetic moment close to 2$\mu_B$, implying the
known diradical nature of O$_2$ in gas phase.
The energetics of adsorption is estimated approximately as $E(O_2^*)-E(^*)-E(O_2)$, where ($^*$) and ($O_2^*$) denote 
bare substrate and $O_2$ adsorbed on substrate respectively. Total energies($E$) are estimated as function of total magnetization. 
Fig.\ref{O2adsorption} shows that adsorption systematically weakens with increasing island size, in agreement with anticipated
lowering of activation of larger islands. Adsorption of O$_2$ on C at B site is consistently stronger than that on C at N sites, 
which is opposite to that we see in case of adsorption of atomic H. 
To understand this we recall that H(O) has lower(higher) electronegativity than C, whereas, C at B site will have in effect 
lower electronegativity than that of a C at N site likely due to back transfer of electron from the lone pairs of N to C at B site.
A more electronegative atom would tend to chemisorb at the less electronegative site and vice-versa, since the strength
of a covalent bond increases with increasing heteropolarity of the bond.
Thus in general all atoms having higher(lesser) electronegative than C would prefer adsorption on C at B(N) site.

\begin{figure*}[t]
\centering
\includegraphics[scale=0.3]{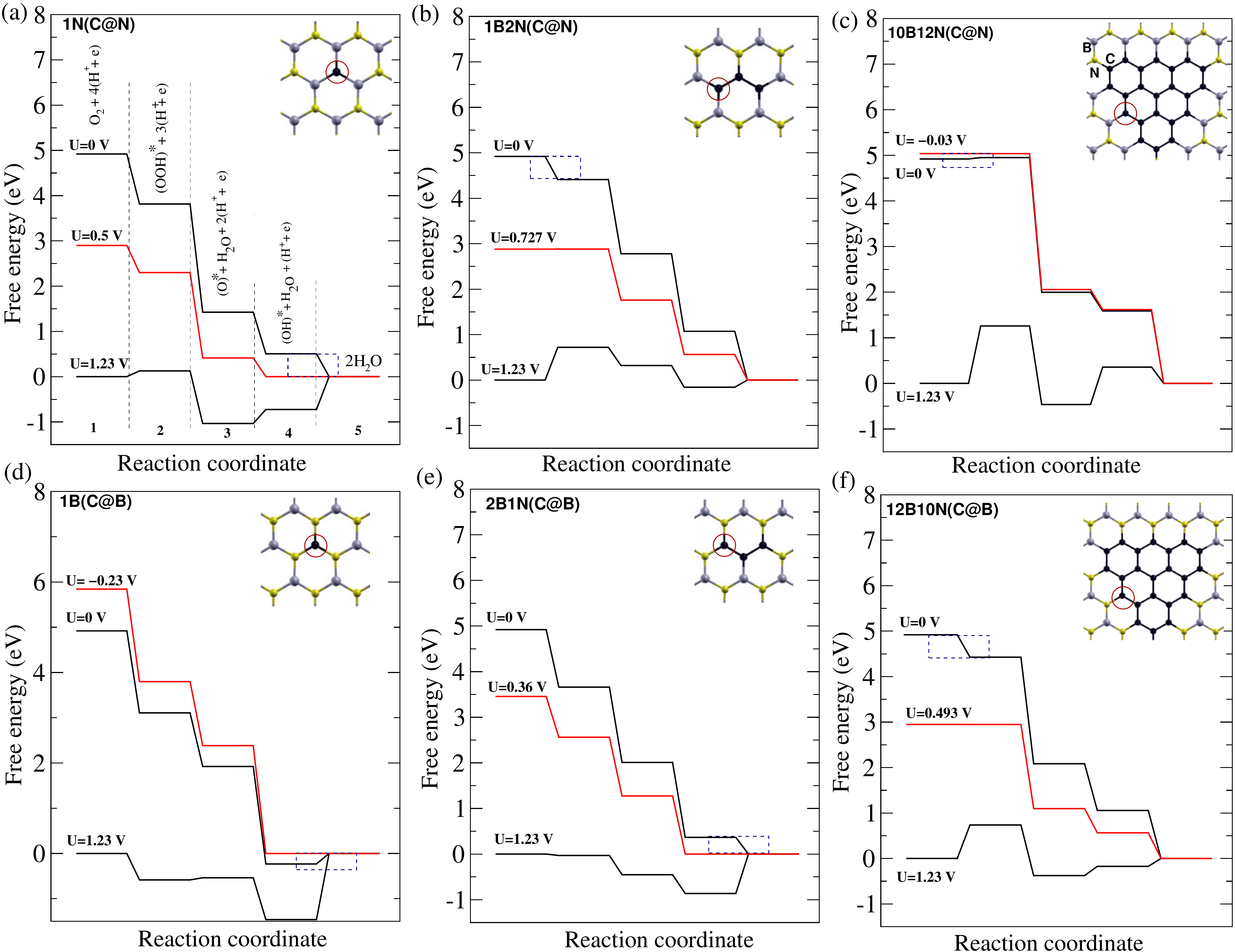}
\caption{Free energy diagram of ORR in acidic condition on different configurations of substitution by C(inset): C at N for (a)1N,
(b)1B2N, (c)10B12N, and C at B for (d)1B, (e)2B1N, (f)12B10N. 
Blue boxes pointing out the rate limiting steps and red circle represents the catalytic site. 
Free energy pathway for maximum operating voltage(U$_{ORR}$) is shown here in red.}
\label{orr}
\end{figure*}

Partitioning of charge upon adsorption of O$_2$ is shown in Fig.\ref{WCO2adsorption} in terms of WCs which are
represented by black translucent spheres. For spin2 in Fig.\ref{WCO2adsorption}(a), the O=O double bond is represented by a 
bigger sphere whose volume is double of that of the smaller spheres representing one electron each.
Spin polarized WC based estimate of charge per atom suggests chemisorption[Fig.\ref{WCO2adsorption}(a)] of O$_2$ 
as a dipole with +0.5e charge on the O attached to the  active site, and -0.5e on the outer O.
dark circle represe
In gas phase O$_2$ exists in the triplet ground-state with a single O-O bond,
implying incomplete sub-shell filling of both O atoms.
Thus, after adsorption on a single active site, the O-O bond order increases from 1 to 1.5, consistent with it's length(1.33\AA{}), 
which is approximately the average of O=O double and O-O single bond lengths.
O$_2$ adsorbs in the same configuration on active B, with similar O-O bond length.      
WCs in Fig.\ref{WCO2adsorption}(b) describe adsorption of O$_2$ as a neutral bridge with O-O single bond on two active sites.  
Free energy of adsorption of O$_2$[Fig.\ref{orr}]
is consistent with our expectation of stronger binding on active B or C at B, than on C at N. 
Notably, energy of adsorption decreases steadily with increasing size of islands due to moderation of
activation of the active sites in larger islands, as argued above.
\subsubsection{Activation of B atoms due to C at B}
Spin separation driven by on-site Coulomb repulsion due to single substitution by C at B, 
as implied by the displacement of WCs shown in Fig.\ref{WC1B1N}(a),
in effect results into back transfer of electron from the nearest neighboring N to the next nearest neighboring B atoms, 
which in effect increases the order of those B-N bonds and renders those B atoms negatively charged.
With excess electrons at their disposal, it becomes easier for those B atoms to chemisorb through a covalent bond or order 1,
an atom in attempt to complete their sub-shell filling.
In Fig.\ref{WCO2adsorption}(b) we indeed see such a scenario, where B chemisorbs O$_2$ and complete sub-shell filling
after receiving an electron from the neighboring N, which thus becomes positively charged, 
but completes sub-shell filling by doubling the order of the bond with the C at B site.
Thus the activation of the C atom gets completely quenched. 
Activation will thus be effective either for the C at B site, or for a B atom in its neighborhood\cite{orr-CatB-hbn-2016}, 
assisted by charge transfer from C to the C-N bond and further from N to B. 
In fact, adsorption of O$_2$ is stronger on such active B, than on the C at B it self, owing to lower electronegativity of B than C. 
Possibility of such activation of B will naturally reduce with increasing distance from the C at B site.
\subsection{Oxygen reduction reaction}
In this section we focus on ORR on magnetic and non-magnetic islands of increasing size.
Complete reduction of O$_2$ to 2H$_2$O at cathode occurs proceeds in two different pathways depending on the competition between
the strength of binding of O$_2$ on the active site and the strength of O-O Bond in the adsorbed configuration.
For O$_2$ adsorbed on active C, whether on a single C atom or as a bridge, reduces through formation of -OOH, 
as shown in the following pathway
[Eq.\ref{orr1a}-\ref{orr1e}].
\begin{subequations}
\begin{align}
* + O_2&=O^*_2, \label{orr1a} \\
O^*_2+ H^+ + e^-&= OOH^*, \label{orr1b} \\ \
OOH^*+ H^+ + e^-&= O^* + H_2O,\label{orr1c} \\
O^* +  H^+ + e^-&= OH^*, \label{orr1d} \\
OH^* +  H^+ + e^-&=  H_2O, \label{orr1e}
\end{align}
\end{subequations}
where `*' denotes the substrate and $X^*$ denotes chemisorbed $X:O_2,OOH,O,OH$.
However, if adsorption of O$_2$ involve an active B site the strong binding of O on B leads to dissociation of O-O bond upon availability of atomic
H, resulting in two adsorbed OH, which can be simultaneously reduced and released as H$_2$O following the steps.
Thus the steps Eq.\ref{orr1b}-\ref{orr1b} would modify to:
\begin{subequations}
\begin{align}
O^*_2+ H^+ + e^-&= OH^*+O^*,  \label{orr2b} \\
OH^*+ O^* + H^+ + e^-&= O^* + H_2O. \label{orr2c} \\
\end{align}
\end{subequations}
Theoretically the  maximum possible voltage output due to reduction of a single O$_2$ molecule is 1.23V \cite{method-ORR}.
But ideally a part of it, known as over-potential($\eta$), is required to  overcome the
uphill steps in course of complete reduction of  O$_2$.
Thus the operating voltage(U), 
available to drive current through the external load, is (1.23-$\eta$)V. 
With cathode at voltage $U$ over anode at steady state, free energy of a reaction coordinate is estimated as a function of $U$ as: 
$G(U) = G(0)-n_e eU$, where $G(0)$ is $G$ calculated at $U=0$.
$G$ is calculated as $E+ZPE - TS$, where $E$ is the total energy and $ZPE$ is the zero point energy calculated as 
$\sum_i\hbar \omega_i$, $\omega_i$ being the frequency of the $i-$th phonon mode\cite{method-qe}. 
Entropy $S$ is considered non-zero only for molecules in gas phase and taken from standard literature\cite{method-std-data}.
T is set to 300K. 
The operating voltage($U$) is given by the minima of free energy difference between successive steps:
$U_{ORR} = \mbox{Min}\left\{\Delta G_i \right\}$,        
where $\Delta G_i= G_i(0)-G_{i-1}(0)$, $i$ denoting the reaction steps mentioned in Eq.\ref{orr1a}-\ref{orr2c}.
The potential which thus remains unavailable to the external load, is the over-potential: 
$\eta_{ORR}= 1.23-U_{ORR} .$
The step($i$) which determines the over-potential is the rate limiting step.  
Free energy steps are thus sensitive to the relative strength of binding\cite{sabatier1,sabatier2}
of the intermediates(-OX: -O, -OOH, -OH) on the catalytic sites. 
\begin{figure}[b]
\begin{center}
\includegraphics[scale=0.15]{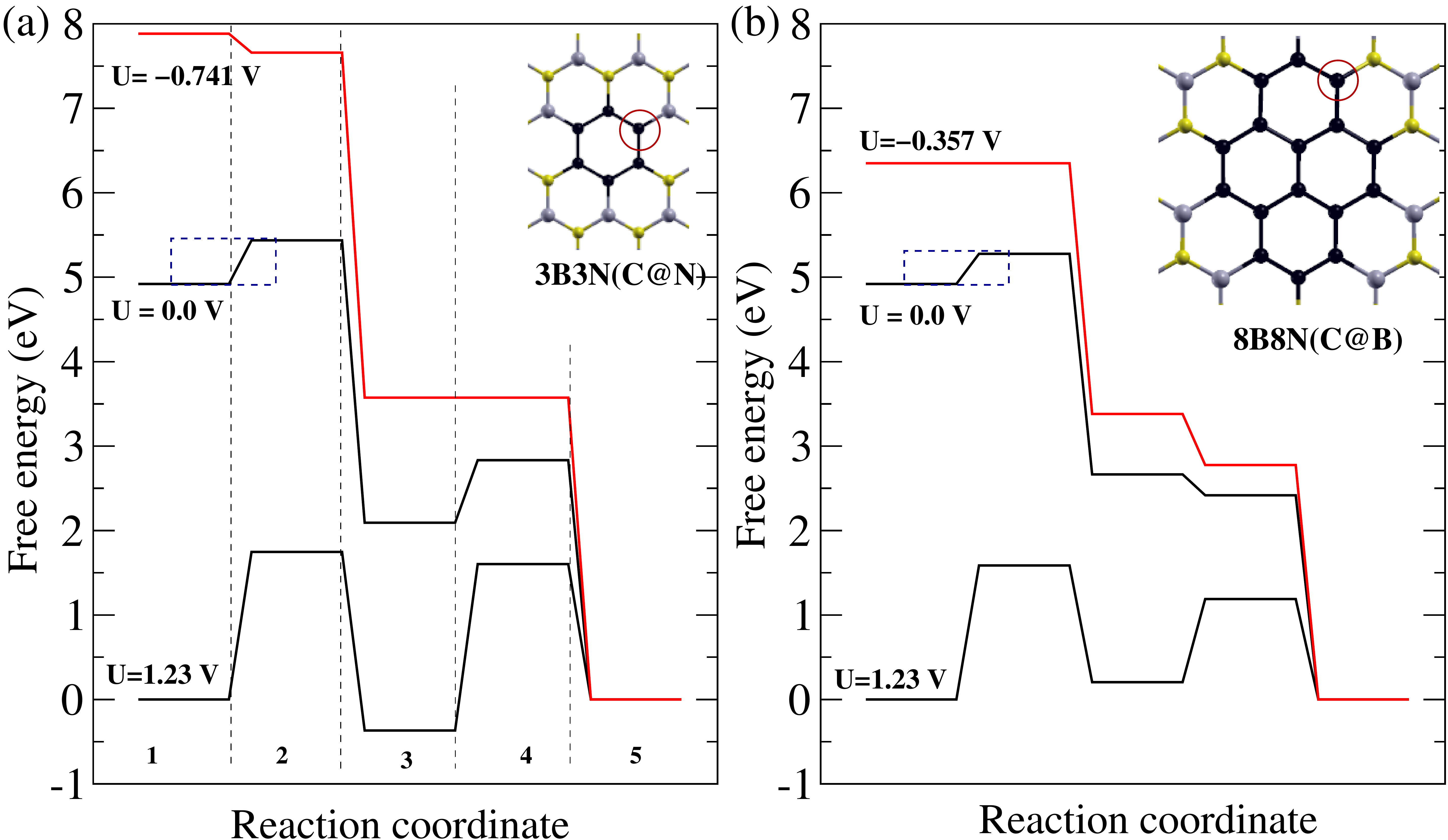}
\caption{Free energy diagram of ORR in acidic condition on non-magnetic configurations of substitution by C(inset) for 
(a) C at N in 3B3N, (b) C at B in 8B8N. 
Dotted boxes pointing out the rate limiting steps and red circle represents the catalytic site. 
Free energy pathway for maximum operating voltage is shown here in red.}
\label{nm_orr}
\end{center}
\end{figure}
\subsubsection{ORR on magnetic graphene island}
As evident from Fig.\ref{orr}(a-c) and Fig.\ref{orr}(d-f), with increasing size of magnetic Gr-islands the rate limiting step
shifts from the last reaction step, which is desorption of O-H,  to the first reaction step which is reduction of O$_2$ to OOH,
since the binding of -OX weakens in general which increases the drop in free energy in the last reaction step.
Fig.\ref{orr}(a,d) shows that modest(strong) binding of -OH on an isolated C at N(B) site leads to workable(negative) operating voltage, 
as has been reported\cite{orr-C-hBN-dft2015}.
Therefore for larger N site rich islands[Fig.\ref{orr}(b,c)], 
weakening of adsorption would lead to lowering of operating voltage, rendering large N site rich islands ineffective for ORR.
On the other hand,  for larger B site rich islands[Fig.\ref{orr}(e,f)] moderation of adsorption would 
lead to increase in of operating voltage, rendering large B site rich islands effective for ORR.
Recalling that B site rich islands are energetically more favorable from, the conclusion that larger B site rich island will be favorable for
ORR allowing workable operating voltages, in effect imply the generic likelihood for C doped hBN to be favorable for ORR.
Solvation energy correction to account for the extra stabilization of -OOH due to accumulation of water molecules surrounding the 
catalytic site has been reported to be less than 0.3V\cite{BG-2014,method-ORR} for doped graphene, which should be applicable in the
present scenario. 
\subsubsection{ORR on non-magnetic graphene island}
Alongside magnetic islands, the non-magnetic Gr-islands, covering an equal number of B and N-sites are also energetically
favorable[Fig.\ref{Eform}]. Here we have analyzed the catalytic activity of two such Gr-islands
(C6 and C16) embedded within hBN. As evident from Fig.\ref{Hadsorption}, all the sites irrespective of C at B or C at N for both C6 and
C16, have comparable activation, which is lower than their counterparts in magnetic Gr-islands.
\begin{figure}[t]
\begin{center}
\includegraphics[scale=0.25]{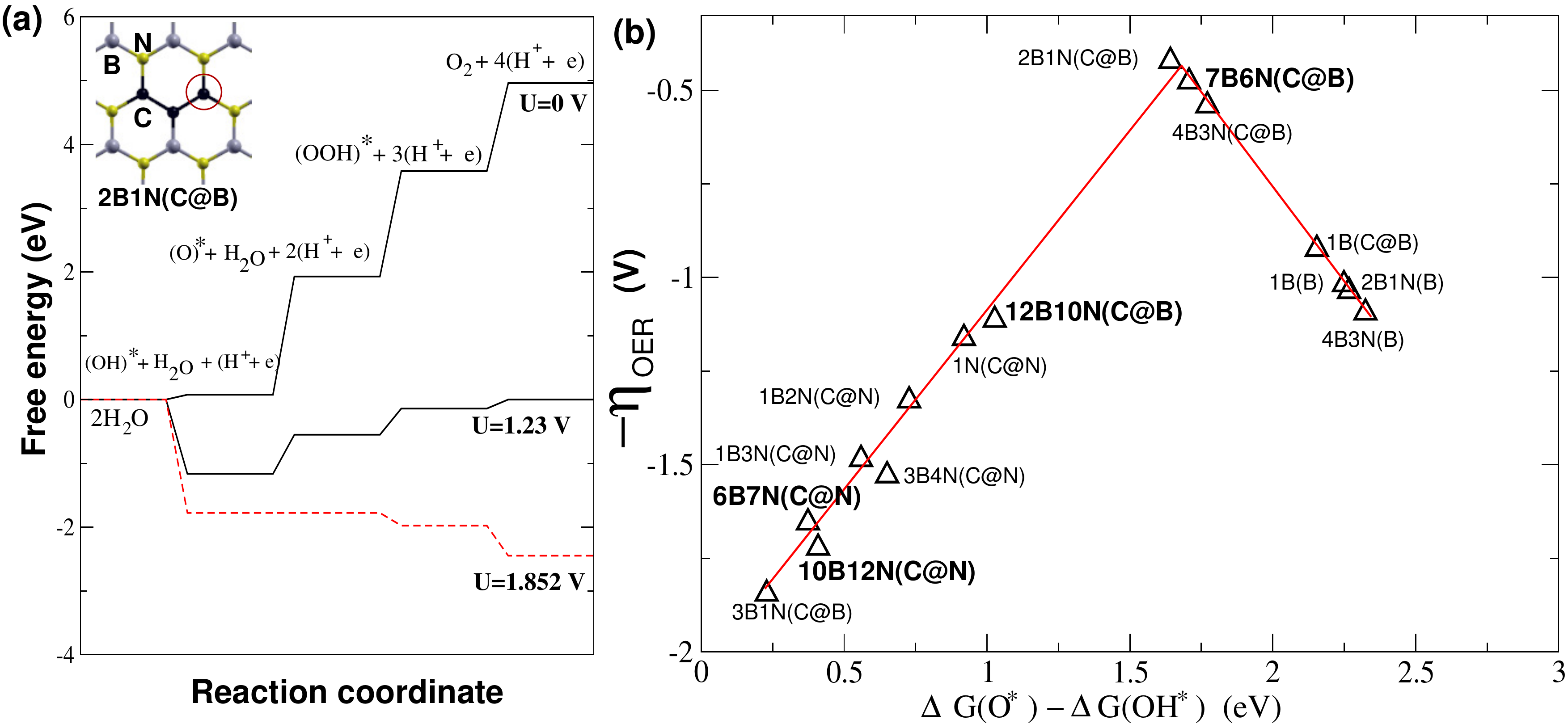}
\caption{(a)Free energy diagram of OER in acidic condition for the configuration with least over-potential and 
(b)Volcano plot for OER w.r.t descriptor $\Delta$G($O^*$)-$\Delta$G($OH^*$)\cite{vol}. 
Red dotted line in (a) corresponds to optimum operating voltage U and the red circle pointing out the
catalytic site. Bigger Gr-islands have been highlighted in bold at plot (b)}
\label{oer}
\end{center}
\end{figure}
Study of ORR reaction pathway on non-magnetic Gr-islands shows an uphill step at the first 
reaction step[Eq.\ref{orr1b}][Fig.\ref{nm_orr}(a,b)]
due to reduced activation of non-magnetic islands.
Therefore the corresponding operating voltage becomes negative, 
which implies high over-potential leading to loss of effectiveness as a cathode material.
\begin{figure*}[t]
\begin{center}
\includegraphics[scale=0.15]{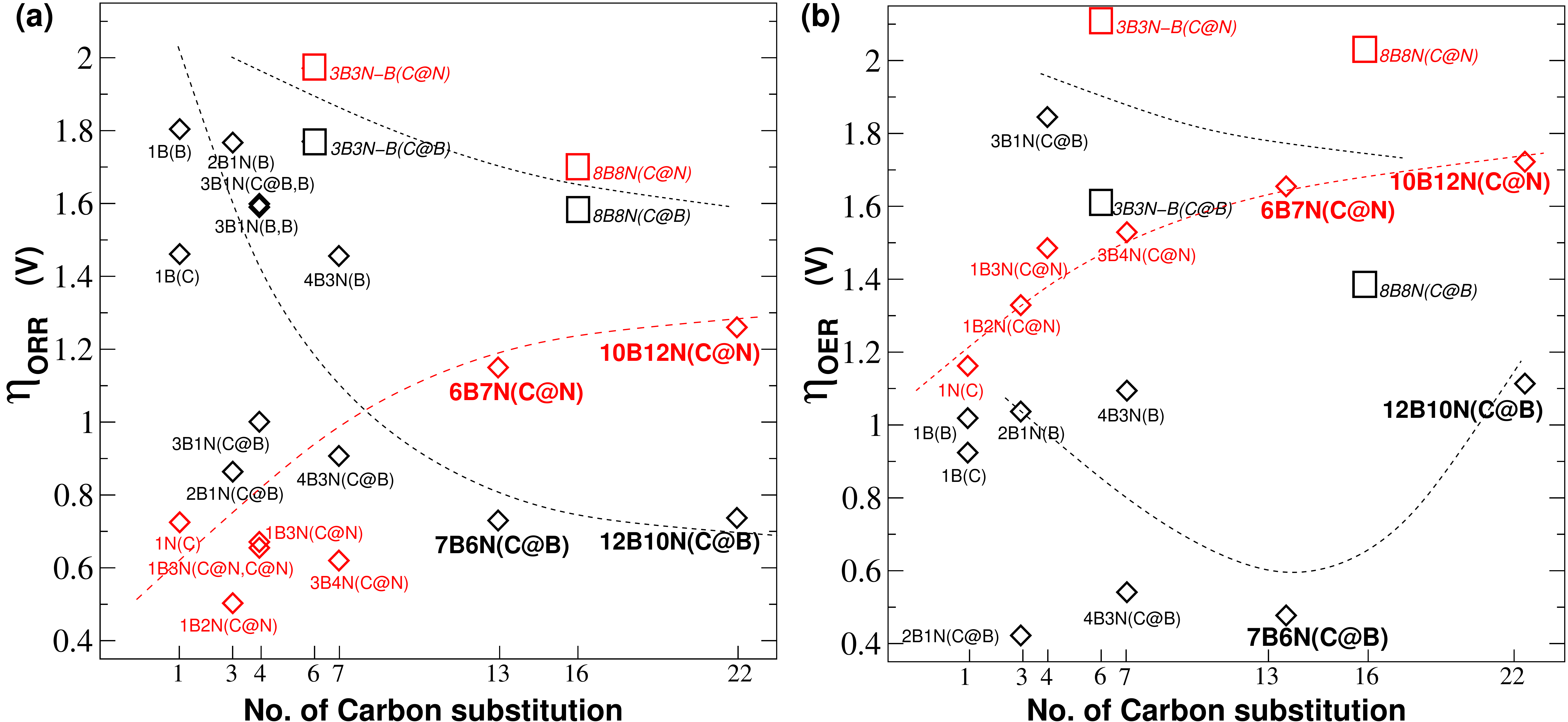}
\caption{Over-potential of all C-substituted hBN configurations (a)for ORR and (b) for OER. 
For magnetic Gr-islands, $\diamond$ in red and black
represent the N-site rich and B-site rich configurations respectively, with the catalytic site mentioned within the parenthesis. 
For non-magnetic islands,
squares on red and black respectively represents catalytic sites C at N and C at B. 
Bigger magnetic islands are highlighted in bold.
The dashed/dotted lines have no significance other than being guides for eye to follow the evolution of over-potential with increasing size
of B(N) site rich magnetic islands (black(red) dashed lines) and non-magnetic islands (black doted line).}
\label{ovp}
\end{center}
\end{figure*}
\subsubsection{OER on magnetic Gr-islands}
Oxygen evolution due to oxidation of water at anode in presence of acidic electrolyte proceeds as [Fig.\ref{oer}(a):
\begin{subequations}
\begin{align}
* + H_2O &=OH^* + H^+ + e^-, \label{oer3a}\\
OH^* &= O^*+ H^+ + e^-,\label{oer3b} \\
O^*+ H_2O &= OOH^*+ H^+ + e^-, \label{oer3c}\\
OOH^* &= O_2 + H^+ + e^-,\label{oer3d}
\end{align}
\end{subequations}
where an O$_2$ is released in gas phase while four H$^+$ and four electrons are passed on to the cathode respectively through the electrolyte and external load.
Conversely to ORR, the operating voltage for OER is given by the maximum difference of successive free energy steps:
$U_{OER} = \mbox{Max}\left\{\Delta G_i \right\}$,        
implying that a over-potential:
$\eta_{OER}= U_{OER}- 1.23$
is essential to drive OER. 
The second step[Eq.\ref{oer3b}] is consistently the rate limiting step for magnetic Gr-islands of different sizes.
Therefore primarily the relative strength of binding of the -O and -OH determine the over-potential.
The volcano plot [Fig.\ref{oer}(b)] drawn accordingly based on Sabatier's principle\cite{sabatier1,sabatier2} suggests 
a clear general trend that active C at B sites are preferred over C at N sites, as catalytic sites for OER, which is expected, 
since stronger binding of -OX intermediates on substrate should make it easier to cleave the O-H bond of water.
However, Fig.\ref{oer}(b) also suggests that the relative binding of -OH over that of -O is preferred to be of about 1.5 eV, 
hinting at a critical level of activation, which is consistent with the
fact that an intermediate size of magnetic graphene island appears to be the most preferable. 
Like ORR, OER[Fig.\ref{ovp}(b)] over-potential is also found high on active sites in non-magnetic Gr-islands.
\begin{figure}[b]
\centering
\includegraphics[scale=0.27]{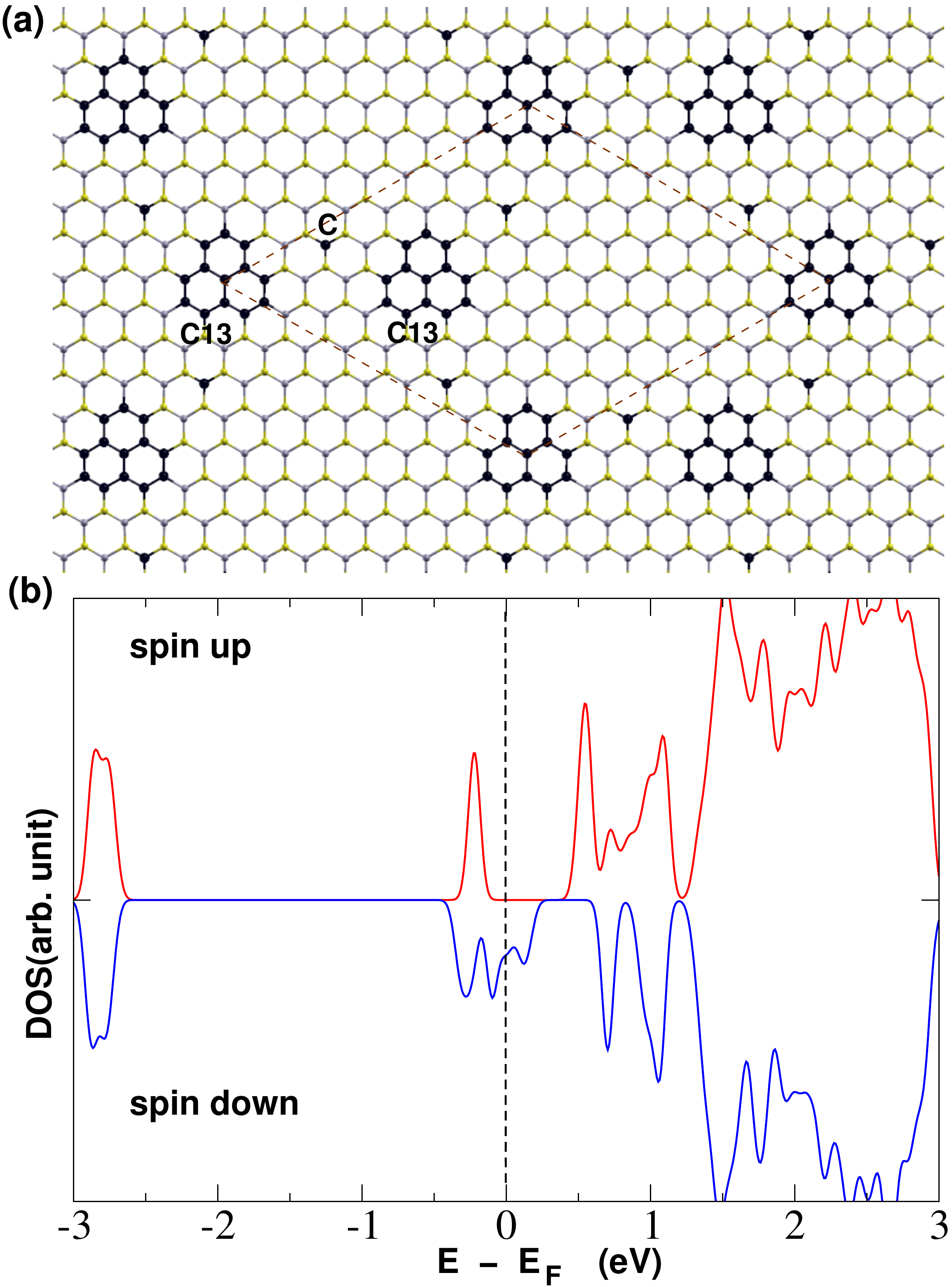}
\caption{(a) Hexagonal-Kagome double-lattice made of C13 and C1 islands respectively (marked in black) in hBN. 
(b) Density of states for the two spins with the Fermi energy set to 0eV. }
\label{HK}
\end{figure}
\subsection{Hybrid super-lattice as cathode material}
The other important aspect which is vital for effective electro-catalysis is transferability of electrons 
from  the leads to the active sites, which is anticipated to be hindered in hBN based systems due to the insulating nature of hBN.
Although metal supported hBN has been proposed\cite{orr-hbn-gold-thr-exp-2014,hbn-metal-2014,rev-orr-hbn-onmetal2015} as a possible
catalytic platform for ORR, our aim is to propose an alternate to metallic electrode exclusively made of non-metals.
In-plane graphene-hBN hybrid has been recently proposed\cite{halfmet} as a half-metal. 
In another recent work\cite{hybrid} of ours we have also extensively investigated the possibility of ferromagnetism and spin polarized transport
in graphene-hBN hybrids, leading to the finding that if the magnetic graphene segments form interpenetrating double lattices
like honeycomb-Kagome [Fig.\ref{HK}(a)], then they can be a half-metal or a ferromagnetic(FM) metal or semiconductor[Fig.\ref{HK}(b)].
C13 and C1 islands, marked in black in Fig.\ref{HK}(a), constitute the honeycomb and Kagome lattices respectively, 
which are ferrimagnetically ordered, leading to a net magnetic moment\cite{hybrid}. Thus the spin degeneracy of the energy bands
will be lifted near the Fermi energy leading to the  half-metallic or FM-metallic or FM-semiconductor phases.
Dispersion of the bands responsible for these phases will increase with increasing size of the magnetic islands, implying
the more robust metallic phases.
\section{Conclusion}
In Fig.\ref{ovp} we summarize the over-potentials for ORR[Fig.\ref{ovp}(a)] and OER[Fig.\ref{ovp}(b)] 
operating voltages for magnetic and non-magnetic islands of different sizes.
Owing to systematic moderation of activation of zigzag interface of graphene islands embedded in hBN, 
the appropriate level of activation of C atoms for efficient ORR is available in 
smaller(larger) N(B) site rich magnetic islands. 
The operating voltage is expected to converged to about 0.5V for B site rich magnetic island of size 1sq-nm or above.
OER is most effective on C at B site at smaller magnetic  islands of size less than 1sq-nm.  
Non-magnetic islands are ineffective for both ORR and OER due to lower levels of chemical activation.
However, since substitution by C in hBN is likely to form magnetic graphene islands almost at par with non-magnetic islands, 
B site rich islands being more likely than their N site rich counterparts, C doped hBN appears to be indeed a promising
platform primarily for ORR catalysis, followed by OER which requires control over island size.
These results thus present a comprehensive view of possibilities for C doped hBN to be used as ORR and OER catalyst.
Thankfully, formation of graphene islands in hBN with controlled shape and size is becoming increasingly possible through recent 
improvements in spatial resolution of irradiation and indentation techniques\cite{doped-ebeam-C-hbn2011,bcn-14}.
We further envisage graphene-hBN hybrid structures with magnetic graphene islands as a possible cathode material exclusively
made of non-metals. 

At a more fundamental level, the work also brings out the central role of on-site Coulomb repulsion in chemical activation 
of not only the C atoms but also of B atoms in the neighborhood, synergistically with magnetism. 
\section{Acknowledgements}
This work has been performed in a high performance computing facility funded by 
Nanomission(SR/NM/NS-1026/2011) of the Dept. of Sci. and Tech. of the Govt. of India.
RM acknowledges financial support from the Dept. of Atomic Energy of the Govt. of India.


\begin{thebibliography}{24}

\bibitem{rev-gra-mat-orr2013}C. Zhu, S. Dong, Recent progress in graphene-based nanomaterials as advanced electrocatalysts towards oxygen reduction reaction,  {\it Nanoscale}, {\bf2013}, 5, 1753-1767.

\bibitem{rev-gra-mat-catal2014}S. Navalon, A. Dhakshinamoorthy, M. Alvaro, H. Garcia, Carbocatalysis by Graphene-Based Materials, {\it Chem. Rev}, {\bf2014}, 114(12), 6179-6212.

\bibitem{rev-B-N-gra-orr2014}Y. Jiao, Y. Zheng, M. Jaroniec, S.Z. Qiao, Origin of the Electrocatalytic Oxygen Reduction Activity of Graphene-Based Catalysts: A Roadmap to Achieve the Best Performance, {\it J. Am. Chem. Soc.}, {\bf2014}, 136, 4394-4403.

\bibitem{orr-N-gra-dft2015}J. Bhattacharjee, Activation of Graphenic Carbon Due to Substitutional Doping by
Nitrogen: Mechanistic Understanding from First Principles, {\it J. Phys. Chem. Lett.}, {\bf2015}, 6(9), 1653-1660.

\bibitem{gnf-oer2016}J. Kang,  Jong-Sung Yu, B. Han, 
First-Principles Design of Graphene-Based Active Catalysts for Oxygen Reduction and Evolution Reactions in the Aprotic Li–O$_2$ Battery, {\it J. Phys. Chem. Lett.}, {\bf 2016}, 7 (14), 2803-2808.

\bibitem{N-gnf-oer2016}HB Yang {\it et al.}, Identification of catalytic sites for oxygen reduction and oxygen evolution in N-doped graphene materials: Development of highly efficient metal-free bifunctional electrocatalyst, {\it Sci Adv.}, {\bf 2016} 2(4): e1501122.

\bibitem{rev-metalfree-orr2015}L. Dai, Y. Xue, L. Qu, H. -J. Choi, J. -B. Baek, Metal-Free Catalysts for Oxygen Reduction Reaction, {\it Chem. Rev.}, {\bf2015}, 115(11), 4823-4892.

\bibitem{orr-C-hBN-dft2015}J. Zhao, Z. Chen, Carbon-Doped Boron Nitride Nanosheet: An Efficient Metal-Free Electrocatalyst for the Oxygen Reduction Reaction, {\it J. Phys. Chem. C}, {\bf2015}, 119 (47), 26348-26354.

\bibitem{orr-gra-hbn-2016}Q. Sun, C. Sun, A. Du, S. Dou, Z. Li, In-plane graphene/boron-nitride heterostructures as an efficient metal-free electrocatalyst for the oxygen reduction reaction, {\it Nanoscale},  {\bf2016}, 8, 14084-14091.

\bibitem{orr-CatB-hbn-2016}M. Gao, M. Adachi, A. Lyalin, T. Taketsugu, Long Range Functionalization of h-BN Monolayer by Carbon Doping, {\it J.Phys.Chem.C}, {\bf2016}, 120(29), 15993-16001.

\bibitem{orr-BCNgra2012}S. Wang, L. Zhang, Z. Xia, A. Roy, D. W. Chang, J. -B. Baek, L. Dai, BCN graphene as efficient metal-free electrocatalyst for the oxygen reduction reaction, {\it Angew. Chem., Int. Ed}, {\bf2012}, 51, 4209-4212.

\bibitem{doped-C-patch-hbn2013-B}SM. Kim,  A. Hsu, P. T. Araujo,  Y. Lee, T. Palacios,  M. Dresselhaus,  J. Idrobo,  K. Kim,  J. Kong, Synthesis of patched or stacked graphene and hBN flakes: a route to hybrid structure discovery. {\it Nano Lett.}, {\bf2013}, 13 (3), 933-941.

\bibitem{syn-gra-hbn-domain2013}Z. Liu, L. Ma, G. Shi, W. Zhou, Y. Gong, S. Lei, X. Yang, J. Zhang, J. Yu, Ken. P. Hackenberg, Ken. A. Babakhani, J. Idrobo, R. Vajtai, J. Lou, P. Ajayan, In-plane heterostructures of graphene and hexagonal boron nitride with controlled domain sizes, {\it  Nature Nanotech.}, {\bf2013}, 8, 119-124.

\bibitem{doped-C-patch-hbn2013-A}S. Tang $et\ al.$, Precisely aligned graphene grown on hexagonal boron nitride by catalyst free chemical vapor deposition, {\it Scientific Reports}, {\bf 2013}, 3, 2666.

\bibitem{hbn-gr-application}Yi Lin, John W Connell, Advances in 2D boron nitride nanostructures: nanosheets, nanoribbons, nanomeshes, and hybrids with graphene, {\it Nanoscale}, {\bf 2012}, 4(22), 6908-6939.

\bibitem{expt-fm-BN-GR}Chong Zhao $et\ al.$, Carbon‐Doped Boron Nitride Nanosheets with Ferromagnetism above Room Temperature, {\it Adv. Funct. Mater.}, {\bf 2014}, 24.


\bibitem{method-dft}W. Kohn, L. J. Sham, Self-Consistent Equations Including Exchange and Correlation Effects, {\it Physical Review.}, {\bf 1965}, 140(4A), A1133-A1138.

\bibitem{method-WF}G. H. Wannier, The Structure of Electronic Excitation Levels in Insulating Crystals, {\it Phys. Rev}, {\bf 1937}, 52, 191.

\bibitem{method-qe}P. Giannozzi, S. Baroni, N. Bonini, M. Calandra, R. Car, C. Cavazzoni, D. Ceresoli, G. L. Chiarotti, M. Cococcioni, I. Dabo, {\it et al.}, 
Journal of Physics: Condensed Matter QUANTUM ESPRESSO: a modular and open-source software project for quantum simulations of materials, {\it J. Phys. Cond. Mat}, {\bf2009}, 21, 395502(1-20).

\bibitem{method-usp}D. Vanderbilt, Soft self-consistent pseudopotentials in a generalized eigenvalue formalism, {\it Phys. Rev. B}, {\bf1990}, 41(R), 7892-7895.

\bibitem{method-pbe}J. P. Perdew, K. Burke, M. Ernzerhof, Generalized Gradient Approximation Made Simple, {\it Phys. Rev. Lett}, {\bf1996}, 77, 3865-3868.

\bibitem{method-bfgs}R. Fletcher, Practical Methods of Optimization, Wiley:  New York; {\bf1987}.

\bibitem{grimme}S. Grimme, Semiempirical GGA‐type density functional constructed with a long‐range dispersion correction, {\it J. Comp. Chem.}, {\bf 2006}, 27, 1787-1799.

\bibitem{mlWF} N. Marzari and D. Vanderbilt, Maximally localized generalized Wannier functions for composite energy bands, {\it Physical Review B}, {\bf 1997}, 56(20).

\bibitem{method-ORR}J. K. N\o{}rskov,  J. Rossmeisl,  A. Logadottir,  L. Lindqvist,  J. R. Kitchin,  T. Bligaard,  H. J \'{o}nsson, Origin
of the Overpotential for Oxygen Reduction at a Fuel-Cell Cathode, {\it J.Phys.Chem.B} {\bf 2004}, 108, 17886-17892.

\bibitem{method-std-data}M. W. Chase Jr., NIST-JANAF Thermochemical Tables, 4th ed., Monograph 9,  {\it J.Phys.Chem}, Ref.Data; NIST: Gaithersburg, MD,{\bf 1998}.

\bibitem{postsynthesis2011}N. Berseneva, A. V. Krasheninnikov, R. M. Nieminen, Mechanisms of postsynthesis doping of boron nitride nanostructures with carbon from first-principles simulations, {\it Phys. Rev. Lett}, {\bf 2011}, 107, 035501.

\bibitem{sabatier1}P. Sabatier, Hydrogénations et déshydrogénations par catalyse, {\it Ber. Deutsch. Chem. Gesellschaft}, {\bf 1911}, 44, 1984.
 
\bibitem{sabatier2}T. Bligaarda $et\ al.$, The Brønsted–Evans–Polanyi relation and the volcano curve in heterogeneous catalysis, {\it Journal of Catalysis}, {\bf 2004}, 224(1).

\bibitem{BG-2014}G. Fazio, L. Ferrighi, C. Di. Valentin, Boron-doped graphene as active electrocatalyst for oxygen reduction reaction at a fuel-cell cathode, {\it Journal of Catalysis}, {\bf 2014}, 318, 203-210.

\bibitem{vol}M. Li $ et\ al.$, N-doped graphene as catalysts for oxygen reduction and oxygen evolution reactions: Theoretical considerations, {\it Journal of Catalysis}, {\bf 2014}, 314.

\bibitem{doped-ebeam-C-hbn2011} X. Wei, M.-S. Wang, Y. Bando, D. Golberg, Electron-beam-induced substitutional carbon doping of boron nitride nanosheets, nanoribbons, and nanotubes, {\it ACS Nano}, {\bf 2011}, 5, 2916-2922.

\bibitem{bcn-14}Y. Gong, $et\ al.$ Direct chemical conversion of graphene to boron-and nitrogen-and carbon-containing atomic layers, {\it Nature Communications}, {\bf 2014}, 5,3193.

 
\bibitem{orr-hbn-gold-thr-exp-2014}K. Uosaki, G. Elumalai, H. Noguchi, T. Masuda, A. Lyalin, A. Nakayama, T. Taketsugu, Boron nitride nanosheet on gold as an electrocatalyst for oxygen reduction reaction: theoretical suggestion and experimental proof., {\it J.Am.Chem.Soc} {\bf2014}, 136(18), 6542-6545.

\bibitem{hbn-metal-2014}G. Elumalai, H. Noguchi, K. Uosaki, Electrocatalytic activity of various types of h-BN for the oxygen reduction reaction, {\it Phys. Chem. Chem. Phys}, {\bf2014}, 16, 13755-13761.

\bibitem{rev-orr-hbn-onmetal2015}R. Koitz, J. K. N\o rskov, F. Studt, A systematic study of metal-supported boron nitride materials for the oxygen reduction reaction, {\it Phys. Chem. Chem. Phys}, {\bf2015} 17, 12722-12727. 

\bibitem{halfmet}Y. Liu, X. Wu, Y. Zhao, X. C. Zeng, J. Yang, Half-metallicity in hybrid graphene/boron nitride nanoribbons with dihydrogenated edges, {\it J. Phys. Chem. C}, {\bf2011} 115, 9442-9450.
\bibitem{hybrid} R. Maji, J Bhattacharjee, Hybrid super-lattices of graphene and hexagonal boron nitride: Ferro-magnetic semiconductor at room temperature, {\it  arXiv:1809.08270v2 } {\bf 2018}.




 
\end{thebibliography}
\end{document}